\documentclass[11pt,english]{article}
\usepackage{jheppub}

\usepackage[T1]{fontenc}
\usepackage[latin9]{inputenc}
\usepackage{amsmath}
\usepackage[toc,page]{appendix}
\usepackage{esint}
\usepackage{babel}
\usepackage{color}
\usepackage{enumitem}
\usepackage{subfigure}

\usepackage{multirow} 
\usepackage{array}
\usepackage{booktabs}
\usepackage{adjustbox}

\usepackage{xargs}
\usepackage[colorinlistoftodos,prependcaption,textsize=tiny]{todonotes}
\usepackage[numbers,sort&compress]{natbib}
\usepackage{hyperref}
\allowdisplaybreaks
\hypersetup{
colorlinks=true,
linkcolor=blue,
linktocpage=true,
citecolor=violet
}

\def\dlog{{$d\log$}}
\def\Mathematica{{\sc Mathematica}}
\def\DlogBasis{{\sc DlogBasis}}
\def\Fire{{\sc Fire6}}
\def\Litered{{\sc LiteRed}}
\def\Finiteflow{{\sc FiniteFlow}}
\def\Polylogtools{{\sc PolyLogTools}}
\def\Ginac{{\sc Ginac}}
\def\pysecdec{{\sc pySecDec}}

\newcommand{\munich}{Max-Planck-Institut f\"ur Physik, Werner-Heisenberg-Institut, 
80805 M\"unchen, Germany.}

\begin{document}
\preprint{MPP-2023-35}

\title{First look at the evaluation of three-loop non-planar Feynman diagrams for Higgs plus jet production}

\author[a]{Johannes M. Henn,}
\author[a]{Jungwon Lim,}
\author[a]{and William J. Torres~Bobadilla}
\affiliation[a]{\munich}
\emailAdd{henn@mpp.mpg.de}
\emailAdd{wonlim@mpp.mpg.de}
\emailAdd{torres@mpp.mpg.de}

\abstract{
We present new computations for Feynman integrals relevant to Higgs plus jet production at three loops, including first results for a non-planar class of integrals. The results are expressed in terms of generalised polylogarithms up to transcendental weight six. We also provide the full canonical differential equations, which allows us to make structural observations on the answer. In particular, we find a counterexample to previously conjectured adjacency relations, for a planar integral of the tennis-court type. Additionally, for a non-planar triple ladder diagram, we find two novel alphabet letters. This information may be useful for future bootstrap approaches.
}

\maketitle
\graphicspath{{figs/}}

\section{Introduction and summary of main results}

Perturbative approaches in quantum field theory are a crucial ingredient for comparing experimental and theoretical predictions. Over the last decades, leading order (LO) and  
next-to-leading order (NLO) calculations have been obtained for many relevant observables, and in some cases even NNLO and NNNLO results are available.
This situation, however, is not the end of story in view of planned upgrades of
the Large Hadron Collider (LHC) that 
involves a high-luminosity phase with 14 TeV center-of-mass energy in proton-proton collisions. With experimental results for scattering processes aiming to reach per-cent level of precision,
new theoretical predictions are required. 

In particular, one of the main interests in high-luminosity phase at LHC is the production of Higgs boson 
in association with jets.
Since the most important mechanism to produce Higgs bosons at LHC is mediated by top quarks, 
one can consider an effective field theory in which the 
top quark mass becomes infinity. Based on this 
effective interaction of gluons and Higgs~\cite{Ellis:1975ap,Shifman:1979eb,Kniehl:1995tn}, 
phenomenological results were provided up to NNLO~\cite{Gehrmann:2000zt, Gehrmann:2001ck,Gehrmann:2011aa,Gehrmann:2023etk}. 

In view of large QCD perturbative corrections for scattering processes that involve Higgs production 
and the constant progress on the experimental side,
higher orders beyond NNLO are crucial.
A major bottleneck for the calculation of NNNLO theoretical predictions 
is obtaining the missing three-loop integrals with one off-shell leg.
The planar three-loop ladder 
integrals~\cite{DiVita:2014pza} were computed some time ago, and more recently the remaining planar integrals, of the tennis-court type, were computed in Refs.~\cite{Canko:2020gqp,Canko:2021xmn}.
However, the three-loop non-planar integrals are not known (with the exception of certain six-propagator integrals~\cite{Henn:2013nsa}).

In the present paper, we compute analytically for the first time a class of three-loop non-planar ladder type diagrams with one off-shell leg. 
We also revisit the planar diagrams, so as to provide all results in a uniform language. 
We construct differential equations by following standard procedures~\cite{Henn:2014qga}, leveraging automated frameworks 
\DlogBasis{}~\cite{Henn:2020lye}, \Litered~\cite{Lee:2012cn},
\Fire{}~\cite{Smirnov:2019qkx}, and \Finiteflow{}~\cite{Peraro:2019svx}. 
We express the analytic solutions of the integrals in terms of generalised polylogarithms (GPLs)~\cite{Gehrmann:2001jv,Goncharov:2010jf} 
up to transcendental weight six.
We numerically evaluate our solutions with \Ginac{}~\cite{Vollinga:2004sn} through \Polylogtools{}~\cite{Duhr:2019tlz}, 
and validate our results via numerical evaluations by \pysecdec{}~\cite{Borowka:2017idc}.

Our analytic results also provide new insights into recent observations on the function space needed for Feynman integrals and form factors \cite{Dixon:2020bbt,Chicherin:2020umh}. These references found in all cases studied in the literature that adjacency conditions hold. At symbol level, this means that certain symbol letters cannot appear next to each other.
Moreover, it was found that the function alphabet is related to a cluster algebra~\cite{Chicherin:2020umh}.
The adjacency relations, as well as parallel developments on integrability \cite{Sever:2020jjx,Sever:2021xga,Sever:2021nsq},
were instrumental bootstrapping three-gluon form factors~\cite{Brandhuber:2012vm} in $\mathcal{N}=4$ sYM to very high loops orders~\cite{Dixon:2020bbt,Dixon:2021tdw,Dixon:2022rse}.

Given these results one might expect that the observed properties hold to high loop orders. Surprisingly, our results show that additional alphabet letters are required for certain loop integrals. 
This means that the letters cannot be all described by the $C_2$ cluster algebra. 
Moreover, by analysing in detail the analytic results for the tennis-court Feynman integrals, 
we find a counterexample to the adjacency relations that had not been noticed previously.

This paper is structured as follows. 
In section~\ref{sec:def}, we set our convention for kinematic configuration and definition 
of integral families considered in this work. 
We discuss the construction and features of differential equations in canonical form for
the integral families in Sec.~\ref{sec:deq}.
We solve the differential equations and provide analytic 
expressions in terms of generalised polylogarithms up to transcendental weight six in Sec.~\ref{sec:sol}. 
Our observations on novel symbol letters and
on adjacency conditions 
are discussed in Sec.~\ref{sec:adj}. 
Finally, in Sec.~\ref{sec:conc}, we draw our conclusions and discuss further research directions.

In the arXiv submission of the current paper, we include ancillary files containing 
information on the computations presented in the next sections.
For each integral family, we provide integral family definition (\verb"family_definition.m"),
set of integrals that satisfy a differential equation in canonical form (\verb"family_can.m"),
canonical matrix (\verb"family_Atilde.m"), and 
analytic solution of integrals in terms of generalised polylogarithms up to 
transcendental weight six (\verb"family_sol.m").

\section{Planar and non-planar integral families for three-loop four-point integrals with one off-shell leg}
\label{sec:def}

\begin{figure}[t]
\centering
\subfigure[Integral family A.]{\label{fig:familyA}\includegraphics[scale=0.9]{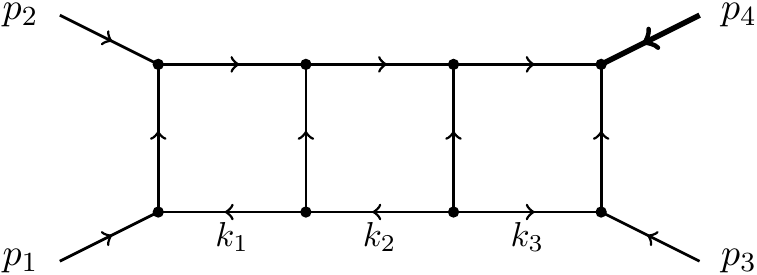}}
\subfigure[Integral families B1 and B2.]{\label{fig:familyB}\includegraphics[scale=0.9]{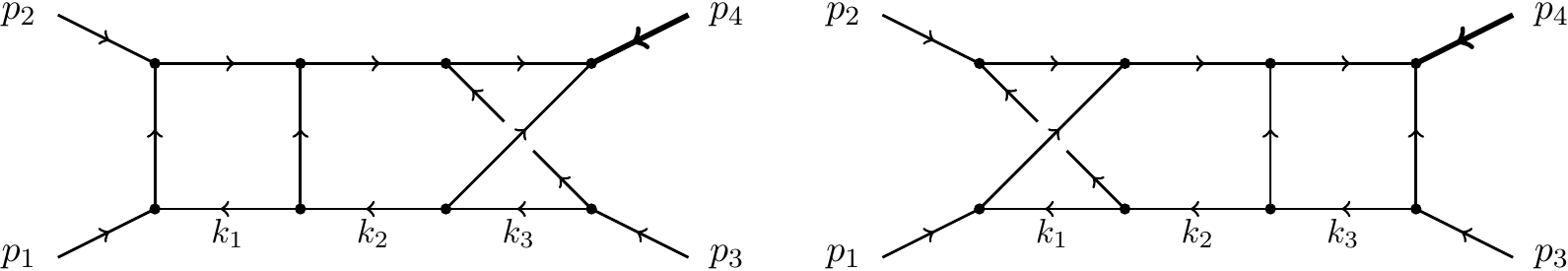}}
\subfigure[Integral families E1 and E2.]{\label{fig:familyE}\includegraphics[scale=0.9]{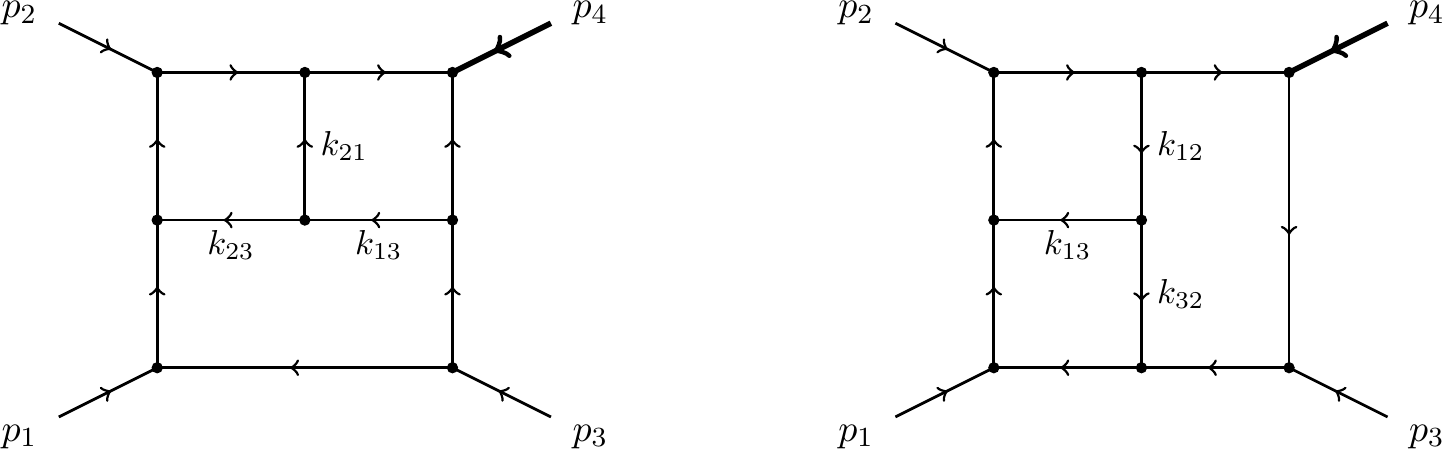}}
\caption{The planar and non-planar integral families considered in this paper. Here $k_{ij}\equiv k_i-k_j$. 
Thin lines indicate one-shell momenta, whilst thick ones indicate off-shell ones.
The labelling of the integral families follows the convention of Ref.~\cite{Henn:2020lye}.
}
\label{fig:families}
\end{figure}

In this section, we introduce the kinematic configuration of external momenta
as well as the convention used throughout this paper for the calculation
of three-loop planar and non-planar Feynman integrals displayed 
in Fig.~\ref{fig:families}. 

\subsection*{Kinematics}
We consider three on-shell ($p_i^2=0$ with $i=1,2,3$) 
and one off-shell external ($p_4^2 \neq 0$) momenta that satisfy momentum conservation,
$p_1+p_2+p_3+p_4=0$, and define the Lorenz invariant scalar products, 
\begin{align}\label{eq:euclidean}
    s = (p_1+p_2)^2\,, \qquad
    t = (p_1+p_3)^2\,, \qquad
    u = (p_2+p_3)^2\,,
\end{align}
with the condition, $s+t+u=p_4^2$, so that only three of them are independent.
We work in the Euclidean region, 
\begin{align}
p_4^2<0\,,\quad s<0 \,,\quad t<0\,,\quad u<0\,,
\label{eq:reg_euclidean}
\end{align}
which means that all results are real-valued expressions.

To regulate divergences, we consider dimensionally regularised Feynman integrals in $D=4-2\epsilon$ space-time dimensions. 
Our expressions for Feynman integrals $J$ are then normalised as follows, 
\begin{equation}
    J_{\text{X};a_1,\hdots,a_{15}}^{\left(L\right)} = (-r_{\Gamma})^{-L}
    \int\prod_{i=1}^{L}\frac{d^{D}k_{i}}{\imath\pi^{D/2}}\frac{1}{\prod_{j=1}^{15}D_{j}^{a_j}}\,,
    \label{eq:J_norm}
\end{equation}
with $\text{X}$ the name of the integral (to be discussed in the next subsection), 
$L=3$ number of loops, $D_i$'s the Feynman propagators that characterised
any Feynman integral, $a_i$'s the exponent of a respective propagator, and, 
\begin{equation}
    r_{\Gamma}=\frac{\Gamma\left(1+\epsilon\right)\Gamma^{2}\left(1-\epsilon\right)}{\Gamma\left(1-2\epsilon\right)}\,.
\end{equation}

\subsection*{Integral families}

In this paper we consider the integral families depicted in Fig.~\ref{fig:families}.
These families consist of a complete set of planar integrals (families A, E1 and E2) 
and two families of non-planar integrals (families B1 and B2).

For instance, the planar integral family  E1 has propagators, 
\begin{align}
 & D_{1}=-(k_{1}-k_{3})^{2}\,, &  & D_{2}=-(k_{1}+p_{1})^{2}\,, &  & D_{3}=-(k_{1}+p_{1}+p_{2})^{2}\,,\nonumber \\
 & D_{4}=-(k_{2}+p_{1}+p_{2})^{2}\,, &  & D_{5}=-(k_{2}-p_{3})^{2}\,, &  & D_{6}=-(k_{2}-k_{3})^{2}\,,\nonumber \\
 & D_{7}=-(k_{1}-k_{2})^{2}\,, &  & D_{8}=-k_{3}^{2}\,, &  & D_{9}=-(k_{3}+p_{1})^{2}\,,\nonumber \\
 & D_{10}=-(k_{3}-p_{3})^{2}\,, &  & D_{11}=-(k_{3}+p_{1}+p_{2})^{2}\,, &  & D_{12}=-(k_{2}+p_{1})^{2}\,,\nonumber \\
 & D_{13}=-(k_{1}-p_{3})^{2}, &  & D_{14}=-k_{1}^{2}\,, &  & D_{15}=-k_{2}^{2}\,,
 \label{eq:famE1}
\end{align}
and the non-planar family B1 has propagators, 
\begin{align}
 & D_{1}=-k_{1}^{2}\, &  & D_{2}=-(k_{1}+p_{1}+p_{2})^{2}\,, &  & D_{3}=-k_{2}^{2}\,,\nonumber \\
 & D_{4}=-(k_{2}+p_{1}+p_{2})^{2}\,, &  & D_{5}=-k_{3}^{2}\,, &  & D_{6}=-(k_{2}-k_{3}+p_{1}+p_{2}+p_{3})^{2}\,,\nonumber \\
 & D_{7}=-(k_{1}+p_{1})^{2}\,, &  & D_{8}=-(k_{1}-k_{2})^{2}\,, &  & D_{9}=-(k_{2}-k_{3})^{2}\,,\nonumber \\
 & D_{10}=-(k_{3}-p_{3})^{2}\,, &  & D_{11}=-(k_{1}-p_{3})^{2}\,, &  & D_{12}=-(k_{2}+p_{1})^{2}\,,\nonumber \\
 & D_{13}=-(k_{2}-p_{3})^{2}\,, &  & D_{14}=-(k_{3}+p_{1})^{2}\,, &  & D_{15}=-(k_{1}-k_{3})^{2}\,.
 \label{eq:famB1}
\end{align}

In integral families~\eqref{eq:famE1} and~\eqref{eq:famB1}, the first ten propagators ($D_i$ with $i=1,2,\hdots,10$)
are understood from the loop topology (see, respectively, Figs.~\ref{fig:familyE} and~\ref{fig:familyB}), 
the remaining five ones ($D_i$ with $i=11,\hdots,15$) are 
 auxiliary propagators that allow, together with the propagators of the loop topology, to express all scalar products 
($k_i\cdot k_j$ and $k_i\cdot p_j$) between loop and external momenta in terms of these propagators. 

The definitions of all integral families displayed in Fig.~\ref{fig:families} are provided 
in the ancillary \Mathematica{} file \verb"family_definition.m" (with family: A,B1,B2,E1, and E2).

\section{System of canonical differential equations for all master integrals}
\label{sec:deq}

In the generation of differential equations for the master integrals 
with respect to the kinematic invariants ($s$, $t$, and $p_4^2$) of the integral families 
presented in the previous section, we rely on the automated codes:
\Litered, \Fire, \DlogBasis, and \Finiteflow.
Since it is known that a good choice of master integrals can significantly lessen complexity 
in their analytic computation, we choose a set of canonical integrals 
(say $\vec{f}_{\text{X}}$ with $\text{X}\in\{\text{A, B1, B2, E1, E2}\}$)
that satisfy an $\epsilon$-factorised form~\cite{Henn:2013pwa}, 
\begin{align}
\partial_\xi \vec{f}_{\text{X}} = \epsilon A_{\text{X};\xi}\,\vec{f}_{\text{X}}\,,
\label{eq:can_deq}
\end{align}
with $\partial_\xi\equiv\frac{\partial}{\partial_\xi}$,  $\xi=s,t,p_4^2$, and 
$A_{\text{X};\xi}$ matrices of each integral X containing rational functions in terms of the kinematic invariants. 

We choose our canonical integrals 
with the help of the \Mathematica{} package \DlogBasis.
Starting from an ansatz of 
integrand, supported by power-counting and absence of ultraviolet singularities, the latter carries out an analysis of iterated residues.

We find the following steps useful in practice. Firstly, depending on the integral family, it may be useful to try different parametrisations 
of the loop momenta.
Secondly, to construct the canonical basis, 
we follow two complementary approaches. 
For integral sectors with up to nine propagators, we use \DlogBasis{} to obtain the canonical basis. For the remaining integral sectors, we construct suitable basis integrals by analysing maximal cuts.
Thirdly, we remark that 
as already pointed out in Ref.~\cite{Henn:2020lye},
to find all independent \dlog{} integrals one may need to include further sectors 
(effectively enlarging the ansatz). 
As in Ref.~\cite{Henn:2020lye}, we prefer to complement the \dlog{} integrals by certain simple UT integrals (e.g. bubble integrals with doubled propagators)~\cite{Flieger:2022xyq}. 
We summarise the construction of canonical basis of each integral family in Table~\ref{table:mis}. 

\begin{table}[t]
\centering
\begin{tabular}{|c|c|c|c|c|}
\hline 
Integral & \# independent & \# additional & \# master integrals & \# letters\tabularnewline
family & \dlog{} integrals & UT integrals & in family & in family \tabularnewline
\hline 
\hline 
A & 75 & 8 & 83 & 7\tabularnewline
\hline 
B1 & 124 & 26 & 150 & 9\tabularnewline
\hline 
B2 & 106 & 8 & 114 & 7\tabularnewline
\hline 
E1 & 151 & 15 & 166 & 7\tabularnewline
\hline 
E2 & 116 & 1 & 117 & 7\tabularnewline
\hline 
\end{tabular}
\caption{Number of \dlog, UT, master integrals integrals, and alphabet letters (see section \ref{sec:deq}) present in each integral family.
}
\label{table:mis}
\end{table}

Once the canonical basis for each integral family is found, we construct their $\epsilon$-factorised
differential equation~\eqref{eq:can_deq}. 
To this end, we generate integration-by-parts (IBP) identities~\cite{Chetyrkin:1981qh,Laporta:2000dsw} 
with the aid of \Litered{} and \Fire{}.
Derivatives of the master integrals with respect to kinematic invariants are computed in an in-house 
\Mathematica{} implementation. 

After having at hand IBPs and canonical basis for each integral family, one is left with combining both 
results to get the differential equation~\eqref{eq:can_deq} -- in particular, the matrices $A_{\text{X};\xi}$. 
Since this operation can be seen as product of (sparse) matrices, evaluating over finite fields our expressions
(to avoid complexity at intermediate steps)
turns out to be a very efficient approach to obtain our differential equations. 
In fact, with the aid of \Finiteflow{} we analytically reconstruct the various matrices $A_\xi$ 
present in~\eqref{eq:can_deq}. 

Finally, with the analytic expressions of $A_\xi$, we are ready to express our canonical differential 
equations~\eqref{eq:can_deq} in terms of the total differential of our canonical master integrals,
\begin{align}
d\vec{f}_\text{X}  = \epsilon\sum_{i=0}^{8}\,\tilde{A}_{\text{X};i}\, d\log\alpha_i\,\vec{f}_\text{X}\,,
\label{eq:deq_can_atildes}
\end{align}
where $\tilde{A}$ are matrices whose entries are rational numbers, and $\alpha$ correspond to letters of the alphabet. 
We find,
\begin{align}
\vec{\alpha}=\left\{ \alpha_{0},\hdots,\alpha_{8}\right\} =&\,
\Big\{ p_4^2,s,t,-p_4^2+s+t,-p_4^2+s,-p_4^2+t,s+t,\notag\\
&\quad-\left(p_4^2-s\right)^{2}+p_4^2t,s^{2}-p_4^2\left(s-t\right)\Big\}\,.
\label{eq:alphabet_st}
\end{align}
In fact, family $\text{A, B2, E1 and E2}$ require only the first seven letters (in agreement with the previous planar results of~\cite{DiVita:2014pza,Canko:2021xmn}), while the full nine-letter alphabet is required for family $\text{B1}$.

For convenience of the reader, we provide as ancillary files \Mathematica{} formatted expressions containing canonical basis ($\vec{f}$) and canonical matrices ($\tilde{A}$) 
for each integral family, respectively, \verb"family_can.m" and \verb"family_Atilde.m",
with family: A, B1, B2, E1, and E2.\\ 

A final remark on the construction of differential equations
in canonical form. 
Since the first letter $\alpha_0=p_4^2$ can be considered as an overall dimension in the normalisation of Feynman integrals,
we can remove it through the change of variables, 
\begin{align}
z_{1}=\frac{-s}{-p_4^2}\,,\qquad z_{2}=\frac{-t}{-p_4^2}\,.
\label{eq:vars_z}
\end{align}
 This effectively amounts to setting $p_4^2=-1$ without loss of generality, which we shall assume in the remainder of this paper.
In this way, we can express the solution of our canonical basis in terms of 
dimensionless variables $z_1$ and $z_2$. 
In these variables, the Euclidean region (\ref{eq:reg_euclidean}) corresponds to 
$0<z_1<1$ and $0<z_2<1-z_1$ 
(or $0<z_2<1$ and $0<z_1<1-z_2$).
The alphabet~\eqref{eq:alphabet_st} in terms of these variables becomes,
\begin{equation}
\left\{ \alpha_{1},\hdots,\alpha_{8}\right\} =\Big\{
z_{1},z_{2},1-z_{1}-z_{2},1-z_{1},1-z_{2},z_{1}+z_{2},
1-2z_{1}+z_{1}^{2}-z_{2},
z_{1}-z_{1}^{2}-z_{2}
\Big\}\,.
\label{eq:alphabet_z1z2}
\end{equation}

In the next sections, we solve the canonical differential equations for each integral family 
in terms of generalised polylogarithms. 
We study the validity of our results by considering various numerical checks.

\section{Explicit solution up to weight six in terms of generalised polylogarithms}
\label{sec:sol}

With the canonical differential equation~\eqref{eq:can_deq}, 
we can naturally express our sets of master integrals as Chen iterated integrals~\cite{Chen:1977oja},\footnote{For the sake of simplifying notation, we drop the subscript ``X'', since this procedure is identically carried out for all integral families studies here.}
\begin{align}\label{eq:Chenformula}
\vec{f}\left(z_{1},z_{2};\epsilon\right)
&=\mathbb{P}\exp\left(\epsilon\int_{\gamma}d\tilde{A}\right)\vec{f}_{0}\left(\epsilon\right)\,,
\end{align}
where $\mathbb{P}$ accounts for the path ordering in the matrix exponential along the contour $\gamma$
in the space of the dimensionless variables $z_1$ and $z_2$ (see Eq.~\eqref{eq:vars_z}),
and $\vec{f}_{0}$ represents the boundary values at the base point of the contour $\gamma$.
In this representation all integrals $\vec{f}$ are given as Laurent expansion in the dimensional parameter $\epsilon$,
\begin{align}
\vec{f}\left(z_{1},z_{2};\epsilon\right) &= \sum_{i=0}^{6} \epsilon^i\,\vec{f}^{(i)}\left(z_{1},z_{2}\right) + \mathcal{O}\left(\epsilon^{7}\right)\,,
\label{eq:f_ep_sol}
\end{align}
and have universal transcendental (UT) degree zero (by considering the degree in $\epsilon^{-n}\to n$). 
For more background material on this topic written in a pedagogical way, see the recent PhD thesis \cite{Zoia:2021zmb}.

Because of the simplicity and linearity of the alphabet in $z_2$, one can easily provide a representation of the integral families in terms of generalised polylogarithms. This can be achieved by properly choosing the contour $\gamma$ or equivalently by integrating one variable at the time, as in Ref.~\cite{Henn:2014lfa}.

We iteratively solve Eq.~\eqref{eq:deq_can_atildes} in terms of the series expansion~\eqref{eq:f_ep_sol},
\begin{align}
\partial_{z_{1}}\vec{f}^{\left(n\right)}\left(z_{1},z_{2}\right) & =A_{z_{1}}\vec{f}^{\left(n-1\right)}\left(z_{1},z_{2}\right)\,,\nonumber \\
\partial_{z_{2}}\vec{f}^{\left(n\right)}\left(z_{1},z_{2}\right) & =A_{z_{2}}\vec{f}^{\left(n-1\right)}\left(z_{1},z_{2}\right)\,,
\end{align}
with $A_{\xi} = \partial_\xi \tilde{A}$ for $\xi=z_1,z_2$. 

Thus, by first integrating over $z_2$, we find the solution up to a function of $z_1$, 
\begin{align}
\vec{f}^{\left(n\right)}\left(z_{1},z_{2}\right)&=\vec{g}^{\left(n\right)}\left(z_{1}\right)+\int_{0}^{z_2}d\bar{z}_{2}\,A_{z_{2}}\left(z_{1},\bar{z}_{2}\right)\vec{f}^{\left(n-1\right)}\left(z_{1},\bar{z}_{2}\right)\,,
\label{eq:int_z2}
\end{align}
that, because of the way how the integration kernel $A_{z_{2}}$ is expressed in terms 
of the letters that display dependence on $z_2$ ($d\bar{z}_{2}/\left(\bar{z}_{2}-b\right)$), 
one can systematically integrate over $\bar{z}_{2}$ by means of generalised polylogarithms (GPLs)~\cite{Goncharov:1998kja}, 
\begin{align}
G\left(\vec{a}_{n};z\right) & \equiv G\left(\vec{a}_{1},\vec{a}_{n-1};z\right)\equiv\int_{0}^{z}\frac{dt}{t-a_{1}}G\left(\vec{a}_{n-1};t\right)\,,\nonumber \\
G\left(\vec{0}_{n};z\right) & \equiv\frac{1}{n!}\log^{n}\left(z\right)\,.
\end{align}
Then, with this solution at hand, we plug it back in the differential equation for $z_1$, 
\begin{align}
\partial_{z_{1}}\vec{g}^{\left(n\right)}\left(z_{1}\right) & =B_{z_{1}}\vec{g}^{\left(n-1\right)}\left(z_{1}\right)\,,
\end{align}
with $B_{z_{1}}$ a matrix whose entries are of the form $1/\left(z_{1}-b\right)$
with $b$ independent of $z_{2}$. 
By explicitly working out this expression, one finds, 
\begin{align}
\vec{g}^{\left(n\right)}\left(z_{1}\right)&=\vec{f}_0^{\left(n\right)}
\notag\\
&\quad+\int_{0}^{z_{1}}d\bar{z}_{1}\left[A_{z_{1}}\left(\bar{z}_{1},z_{2}\right)\vec{g}^{\left(n-1\right)}\left(\bar{z}_{1}\right)-\partial_{\bar{z}_{1}}\int_{0}^{z_{2}}d\bar{z}_{2}\,A_{z_{2}}\left(\bar{z}_{1},\bar{z}_{2}\right)\vec{f}^{\left(n-1\right)}\left(\bar{z}_{1},\bar{z}_{2}\right)\right]\,,
\label{eq:int_z1}
\end{align}
where the integrand inside the squared bracket is independent of the variable $z_2$. 
Since this operation involves taking derivatives on GPLs in which 
the differentiation variable appears in indices as well as in their arguments, 
we profit from \Polylogtools{} built-in functions. 
This solution is finally expressed up to an integration constant $\vec{f}_0^{\left(n\right)}$.
We remark that this procedure effectively amounts to choosing a particular path $\gamma$ in Eq.~\eqref{eq:Chenformula}, namely as the sum of two segments, first along the  horizontal axis, connecting $(0,0)$ to $(z_1,0)$, and second along the vertical axis, connecting $(z_1,0)$ to $(z_1,z_2)$.
The reason we chose this contour of integration for family B1 is that in this way at each step only linear alphabet letters need to be considered.
For families A, B2, E1, E2, 
we chose a different contour, first integrating along the vertical axis, and then along the horizontal axis.

Thus,
our expressions for integrals in terms of unknown constants can 
be expressed as, 
\begin{align}
\vec{f}^{\left(0\right)}&=  \vec{f}_{0}^{\left(0\right)}\nonumber \\
\vec{f}^{\left(1\right)}\left(z_{1},z_{2}\right)&=  M_{z_{1}z_{2}}^{\left(1\right)}\vec{f}_{0}^{\left(0\right)}+\vec{f}_{0}^{\left(1\right)}\nonumber \\
\vec{f}^{\left(2\right)}\left(z_{1},z_{2}\right)&=  M_{z_{1}z_{2}}^{\left(2\right)}\vec{f}_{0}^{\left(0\right)}+M_{z_{1}z_{2}}^{\left(1\right)}\vec{f}_{0}^{\left(1\right)}+\vec{f}_{0}^{\left(2\right)}\nonumber \\
&\ \ \vdots\nonumber \\
\vec{f}^{\left(6\right)}\left(z_{1},z_{2}\right)&=  M_{z_{1}z_{2}}^{\left(6\right)}\vec{f}_{0}^{\left(0\right)}+M_{z_{1}z_{2}}^{\left(5\right)}\vec{f}_{0}^{\left(1\right)}+\hdots+M_{z_{1}z_{2}}^{\left(1\right)}\vec{f}_{0}^{\left(5\right)}+\vec{f}_{0}^{\left(6\right)}\,,
\label{eq:f_ep_sol2}
\end{align}
in which $M_{z_1z_2}^{(n)}$ correspond to 
matrices containing GPLs of transcendental weight $n$,
after taking into account (order-by-order in $\epsilon$) Eqs.~\eqref{eq:int_z2} and~\eqref{eq:int_z1}.
The constants $\vec{f}_0^{(n)}$ are boundary values that are expected to have transcendental weight $n$. 
We find that they can be expressed in terms of multiple zeta values ($\zeta_n$ with $n>1$). 

\begin{figure}[t]
    \centering
    \includegraphics[scale=0.6]{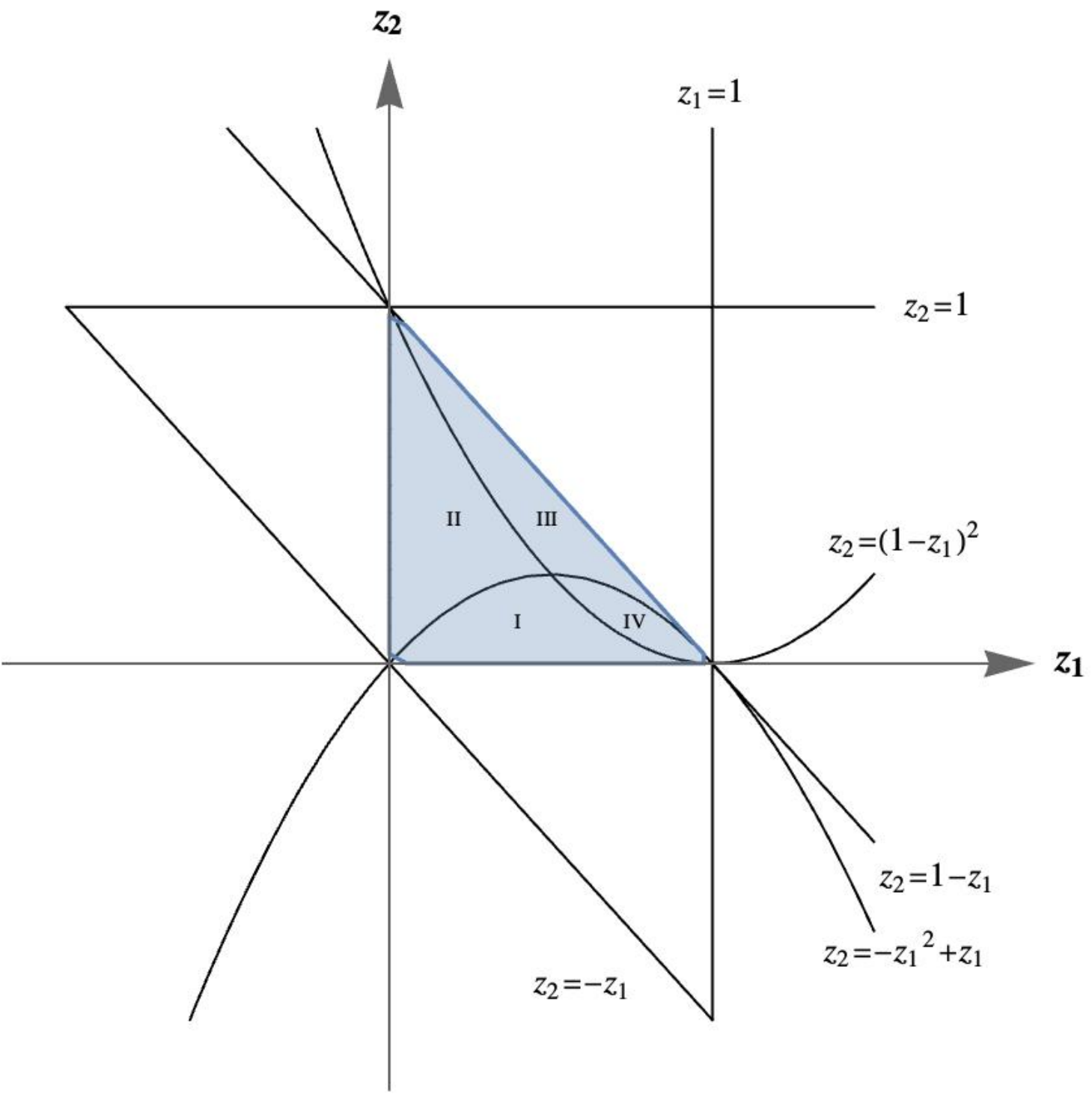}
    \caption{Singular configurations present in the analytic evaluation of our integral families. The shaded region corresponds to the Euclidean region, where all integrals are real-valued. The same is also true for all GPLs
    without dependence on letters $\alpha_7$ and $\alpha_8$, which is the case for integral families A, B2, E1, and E2.
  Our GPL representation for family B1 is manifestly real-valued in region I only (but can be analytically continued to other regions). 
    }
    \label{fig:regions}
\end{figure}

In order to fix the boundary values, we follow the procedure of Ref.~\cite{Henn:2020lye}, 
where we look at all possible (physical and unphysical) threshold singularities that may appear in 
the analytic solution of the integrals. 
This is carried out by investigating all singular limits displayed in Fig.~\ref{fig:regions},
which are given by the letters $\alpha_i\to0$ of alphabet~\eqref{eq:alphabet_st}, 
\begin{align}
\lim_{\alpha_i\to0}\vec{f} &= \alpha_i^{\epsilon\tilde{A}_i}\, \vec{f}(\alpha_i=0)\,,
\label{eq:sing_sol}
\end{align}
with $\vec{f}(\alpha_i=0)$ a vector of boundary constants per each singular limit. 

To extract information from these limits, we bear in mind that our canonical
bases can be chosen to be free from ultraviolet singularities.
Since solutions~\eqref{eq:sing_sol} (with arbitrary boundary vectors) 
may introduce the latter divergences 
when the eigenvalues associated to the matrices $\tilde{A}_i$ are positive, 
we demand that these contributions have to vanish. 

From the constraints imposed on positive eigenvalues, we find linear
relations between boundary values for the various integrals. 
This procedure has been implemented and automated in \Mathematica{} order-by-order in $\epsilon$.
In details, we construct the matrix $\alpha_i^{\epsilon\tilde{A}_i}$, identify the positive eigenvalues 
of this matrix, and evaluate our solutions~\eqref{eq:f_ep_sol2} at the singular limits $\alpha_i=0$
(see Fig.~\ref{fig:regions}). 
We generate, in this way, a set of constraints between boundary constants. 
Once this procedure is performed for all singular limits, we find that all boundary constants
are related to a single one, which sets an overall scale. The latter is computed by direct evaluation. 
In effect, for the calculation of integral families B1 and E1, respectively, 
we only need to consider the analytic expression of the trivial integrals, 
\begin{align}
f^{1}_{\text{B1}}  & = \epsilon^3\,p_4^2\, J_{\text{B1};020000022100000}\,,
\notag\\
f^{1}_{\text{E1}}  & = \epsilon^3\,p_4^2\, J_{\text{E1};002002200100000}\,,
\end{align}
whose analytic expression up to $\mathcal{O}(\epsilon^6)$ is, 
\begin{align}
f^{1}_{\text{B1}} = f^{1}_{\text{E1}} &= 
-1+22\varepsilon^{3}\zeta_{3}+\frac{11\pi^{4}\varepsilon^{4}}{30}+234\varepsilon^{5}\zeta_{5}+\varepsilon^{6}\left(\frac{106\pi^{6}}{189}-242\zeta_{3}^{2}\right)+O\left(\varepsilon^{7}\right)\,.
\end{align}

With all boundary vectors fixed, we can proceed to evaluate our expressions. 
This can be easily done through dedicated routines that numerically evaluate generalised polylogarithms.
For the purpose of presenting results and keeping track of numerical precision, 
we employed \Ginac{} through the interface provided by \Polylogtools. 

We numerically evaluate the analytic expressions of our integrals 
(with precision goal of 30 digits)
in different kinematic points in the Euclidean region,
and validate our results by comparing against the numerical evaluation of the Feynman integrals with \pysecdec{}.
We focus mainly on top sector integrals with simple rank one numerators or without numerators, since these integrals are easier to evaluate for \pysecdec{}.
We do this for all integral families. 
Additionally, we perform dedicated checks for certain integrals in families B1 and E1 that exhibit new features (as will be discussed in Sec.~\ref{sec:adj}).  
We set in \pysecdec{} the precision as relative accuracy $10^{-6}$ for the integrals.
We summarise this comparison in Tables~\ref{table:ints_E1}
and~\ref{table:ints_B1}.

Additionally,
we use the recent program 
\verb"feyntrop"~\cite{Borinsky:2020rqs,Borinsky:2023jdv}, which is based on tropical geometry methods. Using this, we validate all integrals that are expected (from our calculation)
to be finite (which means equivalently that their expansion
starts at $\mathcal{O}(\epsilon^6)$). We find perfect agreement with our GPL results.\\

In ancillary files, we provide \Mathematica{} formatted expressions with the analytic solutions
of integrals for each family, \verb"family_sol.m", with family: A, B1, B2, E1, and E2.

\begin{table}[t]
\centering
\begin{adjustbox}{width=1\textwidth}
\begin{tabular}{|c|c|c|c|c|c|c|c|c|c|}
\hline
 \multirow{2}{*}{Integral} & \multirow{2}{*}{\vbox{\hbox{\strut Evaluation}\hbox{\strut \quad point}}} & \multicolumn{2}{c|}{$\epsilon^3$} & \multicolumn{2}{c|}{$\epsilon^4$} & \multicolumn{2}{c|}{$\epsilon^5$}&  \multicolumn{2}{c|}{$\epsilon^6$}\\ \cline{3-10} 
 &  &  Analytic &\pysecdec{} & Analytic & \pysecdec{} & Analytic & \pysecdec{} & Analytic & \pysecdec{} \\ \hline
\multirow{2}{*}{$f^{110}_{\text{E1}}$} & Point 1 & $0.5045296644$  & $0.504529665(7)$ & $0.7526795133$ & $0.7526794(5)$ & $-0.3064066881$ & $-0.30640(1)$ & $13.8815873594$ & $13.8815(1)$\\ \cline{2-10} 
  &Point 2& $1.3157306457$& $1.3157308(7)$ & $7.0089030292$ & $7.00891(7)$ & $28.4928977317$ & $46.1344(3)$ & $127.7153686313$ & $127.715(1)$  \\ \hline
\multirow{2}{*}{$f^{127}_{\text{E1}}$} & Point 1 & $0.8331985711$ & $0.8331985(1)$ & $3.3361497492$ &$3.33615(1)$ & $11.4545146178$ & $11.45451(6)$ &  $55.7475245548$ & $55.7475(4)$\\ \cline{2-10} 
  & Point 2 & $0.6776332972$ & $0.6776332972(4)$ & $1.2334658424$ & $1.23346584(7)$ & $-1.2713476537$ & $-1.2713475(4)$ & $11.3414720818$ & $11.3414720(9)$ \\ \hline
\end{tabular}
\end{adjustbox}
\caption{Numerical check of integrals $f^{110}_{\text{E1}}$ and $f^{127}_{\text{E1}}$ against \pysecdec{}
at the kinematic points: 
point~1: $\{s,t,p_4^2\} =\{-0.11,-0.73, -1.00\}$, and point~2: $\{s,t,p_4^2\} = \{-0.18, -0.013, -0.25\}$.
}
\label{table:ints_E1}
\end{table}

\begin{table}[t]
\centering
\begin{adjustbox}{width=1\textwidth}
\begin{tabular}{|c|c|c|c|c|c|c|c|c|c|}
\hline
  \multirow{2}{*}{Integral} & \multirow{2}{*}{\vbox{\hbox{\strut Evaluation}\hbox{\strut \quad point}}} & \multicolumn{2}{c|}{$\epsilon^3$} & \multicolumn{2}{c|}{$\epsilon^4$} & \multicolumn{2}{c|}{$\epsilon^5$}&  \multicolumn{2}{c|}{$\epsilon^6$}\\ \cline{3-10} 
 &  &  Analytic &\pysecdec{} & Analytic & \pysecdec{} & Analytic & \pysecdec{} & Analytic & \pysecdec{} \\ \hline
\multirow{2}{*}{$f^{41}_{\text{B1}}$} & Point 1 & $0.3768713705$  & $0.37687137(8)$ & $0.2595847621$ & $0.259585(2)$ & $-24.1653497052$ & $-24.1653(2)$ & $-255.4746048147$ & $-255.474(2)$\\ \cline{2-10} 
  &Point 2& $0.0882252953$& $0.08822531(6)$ & $0.1851070156$ & $0.185107(1)$ & $-3.5650885140$ & $-3.56509(1)$ & $-45.4350139041$ & $-45.4350(2)$  \\ \hline
\multirow{2}{*}{$f^{67}_{\text{B1}}$} & Point 1 & $-6.1800769944$ & $-6.1800771(7)$ & $-37.5823284468$ &$-37.58232(7)$ & $-38.4079844011$ & $-38.4080(4)$ &  $897.7904682990$ & $897.790(7)$\\ \cline{2-10} 
  & Point 2 & $0.3592309958$ & $0.35923099(3)$ & $-1.1083670295$ & $-1.108367(1)$ & $-38.2406764190$ & $-38.2407(1)$ & $-367.9705607540$ & $-367.970(1)$ \\ \hline
\end{tabular}
\end{adjustbox}
\caption{Numerical check of integrals $f^{41}_{\text{B1}}$ and $f^{67}_{\text{B1}}$ against \pysecdec{}
at the kinematic points: point~1: $\{s,t,p_4^2\} =\{-0.11,-0.73, -1.00\}$, and point~2: $\{s,t,p_4^2\} = \{-0.18, -0.013, -0.25\}$.}
\label{table:ints_B1}
\end{table}

\section{New symbol letters and observations on adjacency conditions}
\label{sec:adj}

\subsection{Novel symbol letters in family B1}

\begin{figure}[t]
\centering
\includegraphics[scale=0.9]{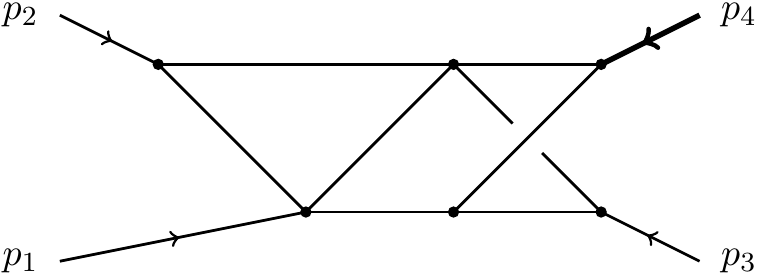}
\quad
\includegraphics[scale=0.9]{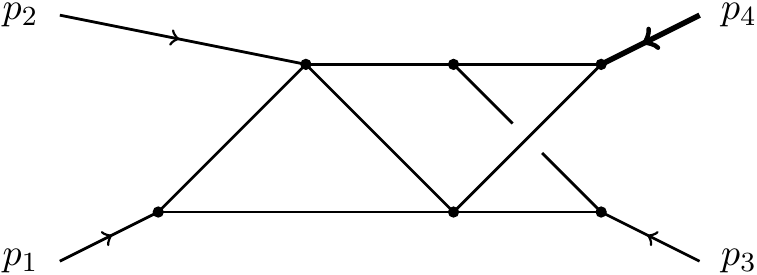}
\caption{
Integrals $f_{\text{B1}}^{41}$ and $f_{\text{B1}}^{67}$
that depend on the letters $\alpha_7$
and $\alpha_8$ of alphabet~\eqref{eq:alphabet_st}, respectively.
The two integrals are related by the symmetry 
$p_1 \leftrightarrow p_2$.}
\label{fig:intnewletters}
\end{figure}

Let us now turn our attention to the new feature of integral family B1, 
namely the two new alphabet letters ($\alpha_{7}$ and $\alpha_8$).
We find that the appearance of new letters are only related to the following integrals (see Fig.~\ref{fig:intnewletters}),
\begin{align}
f^{41}_{\text{B1}}&=\epsilon^6 \left[ \left(p_4^2-s\right)^{2}-p_4^2t\right] J_{\text{B1};011011111100000}\,,
\notag\\
f^{67}_{\text{B1}}&=\epsilon^6 \left[-s^2+p_4^2\left(s-t\right)\right] J_{\text{B1};100111111100000}\,.
\label{eq:intnewletters}
\end{align}
This can be noted by inspecting matrices 
$\tilde{A}_{\text{B1};7}$ and $\tilde{A}_{\text{B1};8}$,
since their matrix rank is one. 

Therefore, a rotation of our complete set of integrals can be performed to only display dependence on new letters in integrals~\eqref{eq:intnewletters}.
Let us illustrate further this statement by considering the integral in top sector (see Fig.~\ref{fig:familyB}), 
\begin{align}
f_{\text{B1}}^{148} & =\epsilon^6 s\left(p_4^2-s\right)^2 J_{\text{B1};11111111110-1000}\,.
\end{align}
From our automatic generation of integrals in canonical form and inspecting $\tilde{A}_{\text{B1};7}$ and $\tilde{A}_{\text{B1};8}$, we realise that this integral is expected to have an explicit dependence on both new letters that, however, can be removed from a rotation, i.e., 
\begin{align}
f_{\text{B1}}^{148} \ \to \
g_{\text{B1}}^{148} = f_{\text{B1}}^{148}
+\frac{1}{3}f_{\text{B1}}^{41}
-\frac{1}{3}f_{\text{B1}}^{67}\,,
\end{align}
with $g_{\text{B1}}$ a new basis in which only two integrals explicitly manifest dependence on the two new letters.

Finally, let us note that the new letters appear for the first time at transcendental weight four.
For example, 
\begin{align}
\mathcal{S}\left(f_{\text{B1}}^{41}\right)\Big|_{\epsilon^{4}}=&6\Bigg[\alpha_{1}\otimes\alpha_{1}\otimes\frac{\alpha_{2}}{\alpha_{4}}\otimes\alpha_{7}-\alpha_{1}\otimes\alpha_{1}\otimes\alpha_{4}\otimes\alpha_{7}+\alpha_{1}\otimes\frac{\alpha_{4}}{\alpha_{2}}\otimes\frac{\alpha_{3}}{\alpha_{1}\alpha_{4}}\otimes\alpha_{7}
\\
&+\alpha_{2}\otimes\alpha_{1}\otimes\frac{\alpha_{1}\alpha_{4}}{\alpha_{3}}\otimes\alpha_{7}+\alpha_{2}\otimes\alpha_{5}\otimes\frac{\alpha_{3}}{\alpha_{1}}\otimes\alpha_{7}-\frac{1}{2}\alpha_{2}\otimes\alpha_{5}\otimes\alpha_{2}\otimes\alpha_{7}+\hdots\Bigg]\,,
\notag
\end{align}
with ellipses corresponding to terms without the letter $\alpha_{7}$, and where we set $p_4^2=-1$ without loss of generality. 

\subsection{Counterexample to adjacency conditions in family E1}

Recently, it was found that
certain Feynman integrals in dimensional regularisation
can be understood in terms of cluster algebras~\cite{Golden:2013xva,Drummond:2019qjk,Chicherin:2020umh,He:2021esx,Henke:2021ity,He:2021non,He:2021eec}.
In particular, Ref.~\cite{Chicherin:2020umh} points out that the alphabet~\eqref{eq:alphabet_z1z2}
can be understood from the $C_2$ cluster algebra. 

Based on available explicit results for 
planar and non-planar one- and two-loop 
Feynman integrals, and the three-loop ladder integral (integral family A of Fig.~\ref{fig:familyA}), it was noticed in Refs.~\cite{Chicherin:2020umh,Dixon:2020bbt} (and conjectured for higher loop Feynman integrals) that the letters  $1-z_i$ and $1-z_j$ for $i\ne j$ never appear next to each other in a symbol.
Analytic results for families E1 and E2 had already been obtained in Ref.~\cite{Canko:2021xmn}, albeit in a form in which checking the adjacency conditions is not straightforward. In fact, only the adjacency condition $\tilde{A}_4\cdot\tilde{A}_6 = 0$ was successfully checked in Ref.~\cite{Canko:2021xmn}.

It is easy to analyse adjacency conditions in the canonical differential equations approach, as we discuss presently.  In our alphabet~\eqref{eq:alphabet_z1z2}, these letters correspond to $\alpha_4,\alpha_5$, and $\alpha_6$.
 The adjacency relations can readily be formulated in terms of the matrices that accompany these letters in the differential equations~\eqref{eq:deq_can_atildes},
i.e., 
\begin{align}
\tilde{A}_i\cdot\tilde{A}_j = 0 \qquad \text{for }i,j\in\{4,5,6\}\qquad\text{with }i\ne j\,.
\label{eq:c2_adj}
\end{align}
For this integral family E1, we confirm that $ \tilde{A}_4\cdot \tilde{A}_6 =  \tilde{A}_6\cdot \tilde{A}_4 =  \tilde{A}_5\cdot \tilde{A}_6 =  \tilde{A}_6\cdot \tilde{A}_5 =0$, as expected. However, we also find that,
\begin{align}
    \tilde{A}_4\cdot \tilde{A}_5
    \ne 0\,, \qquad 
    \tilde{A}_5\cdot \tilde{A}_4\ne 0 \,.
    \label{eq:no_adj}
\end{align} 
This provides a
counterexample to the adjacency relations observed in Refs.~\cite{Chicherin:2020umh,Dixon:2020bbt}.  

\begin{figure}[t]
\centering
\includegraphics[scale=0.9]{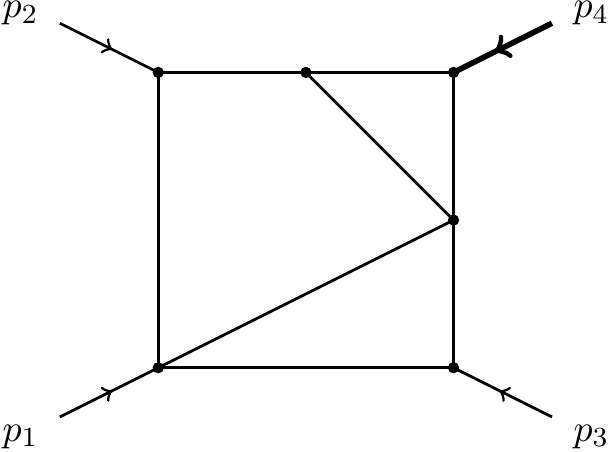}
\quad
\includegraphics[scale=0.9]{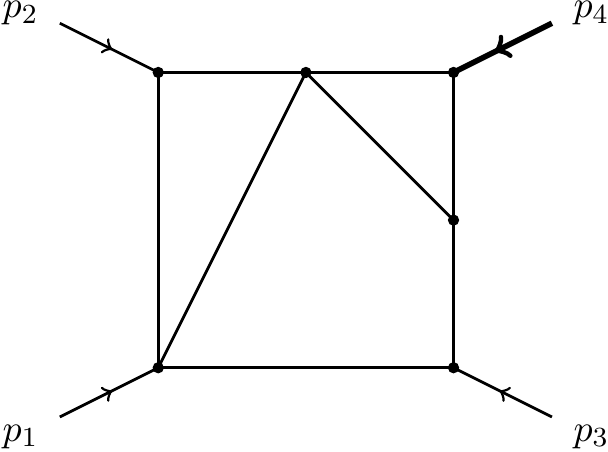}
\caption{Integrals $f^{110}_{\text{E1}}$ and $f^{127}_{\text{E1}}$ 
that violate the adjacency conditions $\tilde{A}_4\cdot \tilde{A}_5=\tilde{A}_5\cdot \tilde{A}_4=0$.
The two integrals are related by the symmetry 
$p_2 \leftrightarrow p_3$.}
\label{fig:intadj}
\end{figure}
Analysing Eqs.~\eqref{eq:no_adj}, we find that the violation of the 
adjacency relations is connected to the following two integrals, 
\begin{align}
    f_{\text{E1}}^{110} & =\epsilon^6 \left(p_4^2-s\right)\left(p_4^2-t\right)J_{\text{E1};111110110100000}\,,
    \notag\\
    f_{\text{E1}}^{127} &=\epsilon^6 \left(p_4^2-s\right)\left(p_4^2-t\right)J_{\text{E1};111111010100000}\,.
\end{align}
The two integrals are related by the symmetry $p_2 \leftrightarrow p_3$,
see Fig.~\ref{fig:intadj}.

The symbol of the solutions is easily obtained from Eq. (\ref{eq:Chenformula}), together with the leading order in $\epsilon$ boundary values.
We find that adjacency-violating symbols in the expressions of $f^{110}_{\text{E1}}$ and $f^{127}_{\text{E1}}$ start appearing at weight five.

\section{Conclusion and outlook}
\label{sec:conc}

In this paper, we calculated all planar, and two non-planar three-loop Feynman integral families with one off-shell leg.
These integrals are relevant, for example, for Higgs plus jet production in the heavy top-quark mass limit of QCD. 
We provided analytic results up to transcendental weight six in terms of generalised polylogarithms, and numerically validated them against \pysecdec{}.

We found that the non-planar integrals we calculated depend on two new alphabet letters that appear for the first time at transcendental weight four. 
Moreover, we studied adjacency relations that had been observed in the literature. We found two counterexamples to these relations, given by the scalar eight-propagator integrals shown in Fig.~\ref{fig:intadj}.
We showed that violation of the adjacency conditions starts at transcendental weight five and six.

There are several interesting directions for further research:
\begin{enumerate} 
\item 
In view of phenomenological applications (e.g. Higgs plus jet production at NNNLO), it would be interesting to compute the remaining non-planar integral families. We expect that obtaining the necessary integral reductions 
could be a bottleneck. However, as we have demonstrated, the ability of predicting a canonical integral basis may streamline this procedure, as significantly reduces the required number of finite fields evaluations.

\item It is interesting to further investigate the function space and adjacency properties, in order to better understand what the counterexamples found in this paper mean. 
What is the reason that form factors in ${\mathcal N=4}$ sYM depend on fewer symbol letters and satisfy adjacency relations? 
What can be said about analogous form factors in QCD? 
Does restricting to four-dimensional finite parts lead to a reduced alphabet, as has been observed in the context of five-particle amplitudes~\cite{Chicherin:2020umh}?
\end{enumerate}

\section*{Acknowledgments}

We thank Simone Zoia for useful discussions.
This research received funding from the European Research Council (ERC) under the European Union's Horizon 2020 research and innovation programme (grant agreement No 725110), {\it Novel structures in scattering amplitudes}, and the Excellence Cluster ORIGINS funded by the Deutsche Forschungsgemeinschaft (DFG, German Research Foundation) under Germany's Excellence Strategy - EXC-2094-390783311.

\newpage

\bibliographystyle{JHEP}
\bibliography{refs}
\end{document}